\documentclass[10pt,a4paper,twoside]{article}
\usepackage{amsmath}
\usepackage{epsfig}
\usepackage{amsfonts}
\usepackage{graphicx}

\newcommand{\ket}[1]{\left | \, #1 \right \rangle}

\def\opone{\leavevmode\hbox{\small1\kern-3.8pt\normalsize1}}

\renewcommand{\vec}[1]{\mbox{\boldmath$#1$}}

\begin{document}

\title{Spin-measurement retrodiction revisited}
\author{Steffen Metzger\thanks{steffen.metzger@physik.uni-muenchen.de}\\
{\em Lady Margaret Hall, University of Oxford, Oxford OX2 6QA, United Kingdom}}
\date{June 2000}
\maketitle

\begin{abstract}

The retrodiction of spin measurements along a set of different axes is revisited in detail.
The problem is presented in two different pictures, a geometric and a general algebraic one.
Explicit measurement operators that allow the retrodiction are given for the case of three and four axes.
For the Vaidman-Aharanov-Albert case of three orthogonal axes the quantum network is constructed for two different initial Bell states.

\end{abstract}

\bigskip
\tableofcontents
\newpage

\section{Introduction}

In contrast to classical mechanics, quantum mechanics does not
allow all observables of a physical system to be measured
simultaneously. Only for a set of mutually commuting observables
definite values can be attributed to a physical system. Therefore
it is one of the predictions of quantum mechanics that two spin
components of a spin-$1\over2$ particle cannot be measured
simultaneously. That is, a given spin state cannot be an
eigenstate of two or more spin operators $\vec{\hat \sigma} \cdot
\vec n_l$, with the usual Pauli matrices $\hat {\vec\sigma} =
(\hat\sigma_x,\;\hat\sigma_y,\;\hat\sigma_z)$ and unit vectors
$\vec n_l$. The result of a future measurement on a state can
only be known in advance if the state is an eigenstate of the
measurement operator. Now, if the spin is to be measured along a
set of axes $\{\vec n_l\}$, then at most the result of one of the
measurements can be predicted, namely when the spin state is an
eigenstate of $\vec{\hat\sigma}\cdot\vec n_i$ for a specific $i$.

However, we can ask the question, whether it is possible to
\textit{retrodict} the results of a spin measurement along more
than one possible axis. To be more specific we have the following
problem: Alice prepares an initial spin-$1\over2$ particle and
gives it to Bob. Bob now performs one spin measurement along one
of a set of axes the directions of which are known to Alice.
However, Alice does not know which axis out of the set Bob
chooses. After his measurement Bob gives the spin-$1\over2$
particle back to Alice. Now Alice can perform another measurement
on the particle, which should enable her to retrodict the value
Bob obtained in his measurement. The fact that Alice does not yet
know the axis along which Bob actually measures is most
important. It means that she must be able to infer all possible
results as a function of the axis that was chosen by Bob. After
her measurement Bob tells her along which axis he actually
measured and she has to tell him the result he got with
certainty. This means that Bob can even cheat and tell Alice that
he measured along axis 1 and Alice tells him which result (she
believes!) he got. But then Bob confesses that he had lied and in
fact measured along axis 2 and again Alice should be able to tell
Bob the result he obtained if indeed he had measured along axis
2. And now Bob might have lied again and measured along another
axis, and so on. This means that the measurement Alice makes has
to extract information about all possible measurement axes out of
the system, even though only one measurement was performed. The
question now is: how should Alice prepare the initial
spin-${1\over2}$ particle and what measurement does she have to
perform on the particle Bob gives back to her?

To state the problem more clearly we put it into mathematical language.
Alice prepares a spin-$1\over2$ particle in a (possibly entangled) state $\vert\psi\rangle_{AB}\in
\mathcal{H}_{AB}:=\mathcal{H}_A\otimes\mathcal{H}_B$. $\mathcal{H}_B$
is the two-dimensional Hilbert space of the spin-$1\over 2$ particle on which Bob can perform a spin measurement and
$\mathcal{H}_A$ is an arbitrary Hilbert space.  The state $\vert\psi\rangle_{AB}$ is measured by Bob along an axis
$\vec n_l \in \lbrace \vec
n_l : 1 \leq l \leq m\rbrace$, i.e. Bob applies the measurement operator
$\opone\otimes(\hat {\vec \sigma} \cdot \vec n_l)$. This measurement projects
the original state onto an eigenstate
\begin{equation}
    \vert\phi_{\eta_l}(\vec n_l)\rangle={1\over 2}(\opone\otimes\opone+\eta_l\opone\otimes(\hat
    {\vec\sigma}\cdot \vec n_l))\vert\psi\rangle_{AB}   \label{one}
\end{equation}
of the spin
operator with eigenvalue $\eta_l=\pm1$. Note that these states are not
normalized if we assume normalisation for the initial state $\vert\psi\rangle_{AB}$.  Now Alice can perform a measurement on
$\vert\phi_{\eta_l}(\vec n_l)\rangle$ and afterwards has to know the
result $\eta_l$ Bob obtained,
along whatever axis he measured. We denote the operator Alice applies
to project onto a basis by $\widehat{\mathcal{M}}=\sum_j
\lambda_j\vert\phi_j\rangle\langle\phi_j\vert$ with
$\lbrace\vert\phi_j\rangle\rbrace$ a basis of  $\mathcal{H}_{AB}$.
The task is to find  $\vert\psi\rangle_{AB}$ and $\widehat{\mathcal{M}}$.
\newpage
The whole process can be visualized like this\footnote{The fact that
  Alice `sends' her state to Bob should not be taken literally. She can as well prepare a state, leave and then Bob comes
  and looks at the state. Then spin states do not have to be parallely
  transported and thus complications can be avoided.}:

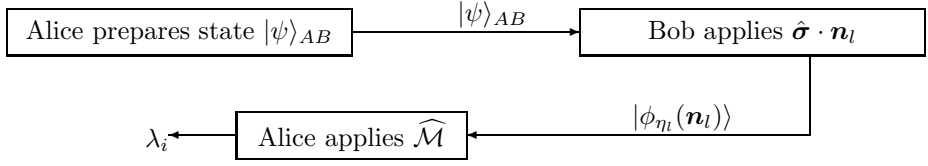
\setlength{\unitlength}{0.030in}
\mbox{\hspace{3cm}}
\mbox{
\begin{picture}(30,30)(-30,15)
  \put(-44,34){\framebox(60,8){Alice prepares state $\vert\psi\rangle_{AB}$}}
  \put(33,40){ $\vert\psi\rangle_{AB}$}
  \put(56,34){\framebox(60,8){Bob applies $\hat{\vec\sigma}\cdot\vec n_l$}}
  \put(16,38){\vector(1,0){40}}
  \put(96,20){\vector(-1,0){60}}
  \put(96,20){\line(0,1){14}}
  \put(65,22){$\vert\phi_{\eta_l}(\vec n_l)\rangle$}
  \put(-4,16){\framebox(40,8){Alice applies $\widehat{\mathcal{M}}$}}
  \put(-4,20){\vector(-1,0){12}}
  \put(-20,18){$\lambda_i$}
\end{picture}}

Let us illustrate the problem by looking at the simple examples
in which $m$=1 and $m$=2, where $\vert\psi\rangle_{AB}$ is unentangled and
$\mathcal{H}_{A}$ is trivial. Therefore $\opone\otimes(\hat {\vec\sigma} \cdot\vec n_l)$ reduces to $\hat {\vec\sigma} \cdot\vec n_l$. The $m$=1 case is trivially solved by either
Alice preparing $\vert\psi\rangle_{AB}$ as an eigenstate of $\hat
{\vec\sigma} \cdot \vec n_1$ or by Alice measuring along $\vec n_1$ afterwards. In the
first case the result is not even retrodicted but can be predicted.

The $m$=2 can be solved easily as well. If Alice prepares an
eigenstate of $\hat{\vec\sigma} \cdot\vec n_1$, she knows in
advance which result Bob gets if he applies
$\hat{\vec\sigma}\cdot\vec n_1$. To infer the value Bob obtains
if he measures along $\vec n_2$, Alice applies the measurement
operator $\hat{\vec\sigma}\cdot\vec n_2$ on the state Bob gave
back to her. This procedure enables her to tell the value Bob
obtained along whatever axis he measured. In fact, here we have a
mixture of a prediction and a retrodiction, as the result of one
possible measurement is known from the beginning.

At first glance the knowledge of spin components along different
axes seems to contradict quantum mechanics. This led Vaidman,
Aharanov and Albert (VAA) who first stated and solved the problem
for three orthogonal axes to use the provocative title ``How to
Ascertain the Values of $\hat\sigma_x,\;\hat\sigma_y$ and
$\hat\sigma_z$ of a Spin-$1\over2$ Particle''  \cite{Aharanov}.
But a closer examination of the problem reveals that the kind of
information Alice has about different spin components is not the
one forbidden by quantum mechanics. Alice, of course, does not
really know two or more eigenvalues of non-commuting operators
but only extracts conditional information. The values Alice gets
are a function of Bob's measurement axis and not until she knows
along which axis Bob actually measured do these values have any
physical meaning. Only when the additional information about the
choice of axis is revealed, \textit{one} value is singled out.
This one now does have physical significance, namely it is the
one eigenvalue Bob actually obtained. This also points out that
if Bob cheats and tells Alice that he had measured along an axis
different from the one he actually used, the result Alice tells
him is meaningless.

\medskip
Obviously, the Hilbert space $\mathcal{H}_A$ was not used in the
$m$=1,2 examples given above and it was sufficient to prepare an
unentangled state in the two-dimensional space $\mathcal{H}_B$. A
state in a two-dimensional Hilbert space is usually called a
\textit{qubit} and is written as $a\ket\uparrow+b\ket\downarrow$,
where $\ket\uparrow$ and $\ket\downarrow$ denote an orthonormal
basis. (For spin-$1\over2$ particles these states are often
eigenstates of spin measurements along a specific axis. E.g. for
the z-axis they are denoted by  ${\ket\uparrow}_z$ and
${\ket\downarrow}_z$.)

The problem no longer has a simple solution if Bob can measure
along three or more axes, which may or may not be orthogonal. Let
us look at the case where Bob can measure along the three
orthogonal axes $x$, $y$ and $z$, i.e. one of the measurement
operators
$\opone\otimes\hat\sigma_x,\opone\otimes\hat\sigma_y,\opone\otimes\hat\sigma_z$
is applied. We can convince ourselves that it is no longer
sufficient to prepare an unentangled qubit if $m\geq$ 3. Hence, if
the problem can be solved at all, $\mathcal{H}_A$ can no longer
be trivial.

The problem as stated above was first presented by VAA in \cite{Aharanov}, who
solved it for the $m$=3 orthogonal case.
Alice's initial state was prepared as
\begin{equation}
    {\ket\psi}_{AB}={1\over\sqrt{2}}\left({\ket\uparrow}_z\otimes{\ket\uparrow}_z+
    {\ket\downarrow}_z\otimes{\ket\downarrow}_z\right) \label{fifteen}
\end{equation}
and Alice's measurement operator ${\mathcal{\widehat M}}=\sum_j \lambda_j \ket{\phi_j}\langle\phi_j\vert$ with $\ket{\phi_j}$
as given in the appendix. In the next section a similar solution for the case that
$\vert\psi\rangle_{AB}$ is the singlet state will be derived.

In \cite{Ben-Menahem} Ben-Menahem used a more algebraic way to
generalise the problem and showed that there are solutions for
the cases $m$=3 and $m$=4, but not for $m\geq5$. However for
$m$=4, the four axes Bob may measure along are no longer
independent but have to satisfy the condition $\sum_{l=1}^{4} \vec
n_l=0$ for there to be a solution. In this case, the dimension of
the Hilbertspace $\mathcal{H}_{AB}$ is shown to be six. This
space, however, is not sufficient to solve the $m$=3
non-orthogonal case, where $\mathcal{H}_{AB}$ has to be
eight-dimensional.

Very recently, this method of spin-measurement retrodiction was
used to set up a secure key distribution in quantum cryptography
\cite{Bub}.

\section{Solution for three orthogonal axes}

After having stated the problem and given some simple examples
with $m$=1 and $m$=2, we can now try to construct a solution for
higher dimensional cases. For the moment we restrict ourselves to
the case in which Bob can measure along three orthogonal axes,
that is either of the measurement operators
$\opone\otimes\hat\sigma_x,\opone\otimes\hat\sigma_y,\opone\otimes\hat\sigma_z$
can be applied to the original state  $\vert\psi\rangle_{AB}$.
Each of the measurements has two possible outcomes $\eta_l=\pm1$,
so there are six possible results altogether, which are denoted by
$\vert\phi_{\eta_l}(\vec
n_l)\rangle\in\lbrace\vert\uparrow,x\rangle$,
$\vert\downarrow,x\rangle$, $\vert\uparrow,y\rangle$,
$\vert\downarrow,y\rangle$, $\vert\uparrow,z\rangle$,
$\vert\downarrow,z\rangle\rbrace\subset\mathcal{H}_{AB}$. Note
that these are different from the basis states of the
two-dimensional Hilbert space $\mathcal{H}_B$ which are denoted
by $\lbrace{\ket\uparrow}_x$, ${\ket\downarrow}_x\rbrace$,
$\lbrace{\ket\uparrow}_y$, ${\ket\downarrow}_y\rbrace$ or
$\lbrace\vert\uparrow\rangle_z$,$\vert\downarrow\rangle_z\rbrace$.
They will only coincide if the original state
$\vert\psi\rangle_{AB}$ is an unentangled qubit.\footnote{Even
then they will only be isomorphic, as
  ${\ket\uparrow}_x\in\mathcal{H}_B$ and $\vert\uparrow,x\rangle\in\mathcal{H}_{AB}$ with $\mathcal{H}_A$  trivial.}

Now Alice measures one of these states $\vert\phi_{\eta_l}(\vec n_l)\rangle$, i.e. she applies the operator
$\widehat{\mathcal{M}}$ to it and therefore projects the state onto a basis
$\lbrace\vert\phi_j\rangle\rbrace$ of $\mathcal{H}_{AB}$ with distinct
eigenvalues  $\lambda_j$ respectively.
To ensure that her measurement gives the result she wants, Alice has to
construct a look-up table. This table must tell her that if she measures e.g. $\lambda_2$ Bob got `up' if he measured
along x, `up' if he measured along y and `down' if he measured along
z. That is, Alice needs a table like this:

\bigskip
\begin{center}
\begin{tabular}{|c||c|c|c|} \hline
         &     x  &     y  &     z     \\  \hline\hline
$\lambda_1$&$\;\;\;\;\downarrow\;\;\;\;$      & $\;\;\;\;\downarrow\;\;\;\;$&$\;\;\;\downarrow \;\;\;\;$\\  \hline
$\lambda_2$&$\uparrow$&$\uparrow$&$\downarrow$ \\  \hline
$\lambda_3$&$\downarrow$&$\uparrow$&$\uparrow$ \\  \hline
$\lambda_4$&$\uparrow$&$\downarrow$&$\uparrow$ \\  \hline
\end{tabular}
\end{center}
\bigskip

So if she measures $\lambda_2$, for example, she immediately knows
that Bob sent her one of $\lbrace \ket{\uparrow,x},
     \ket{\uparrow,y},
     \ket{\downarrow,z} \rbrace$. If Bob now tells her the axis along
     which he measured, Alice is able to tell which result he got.
This leads to a geometric way of solving the problem by
visualizing the  $\vert\phi_{\eta_l}(\vec n_l)\rangle$ and
$\ket{\phi_j}$ states as vectors in $\mathcal{H}_{AB}$. Alice
projects the vector she gets from Bob onto her basis
$\lbrace\vert\phi_j\rangle\rbrace$ and gets a result $\lambda_i$.
But then she has only enough information if each of her basis
vectors is orthogonal to three of the six possible vectors Bob
can send. Only then can she exclude these three and only the
three other vectors can lead to the specific $\lambda_i$ she
measured. If she chooses the states orthogonal to each of her
basis vectors to be one of each of the three pairs $\lbrace
\lbrace\ket{\uparrow,x}, \ket{\downarrow,x}\rbrace,
\lbrace\ket{\uparrow,y}, \ket{\downarrow,y}\rbrace,
\lbrace\ket{\uparrow,z}, \ket{\downarrow,z}\rbrace \rbrace$ the
problem is now solved. If Bob tells her the axis along which he
measured, Alice knows the state he obtained.

This geometric point of view yields the defining equations for the
basis $\lbrace\vert\phi_j\rangle\rbrace$ for a given
$\vert\psi\rangle_{AB}$. As mentioned above, Alice cannot send an
unentangled qubit to solve the problem. This suggests preparing
$\vert\psi\rangle_{AB}$ as an entangled state of two qubits
yielding a four-dimensional Hilbert space $\mathcal{H}_{AB}$. The
entanglement of the states is necessary as Bob will measure only
in a two-dimensional subspace of $\mathcal{H}_{AB}$. This means
that, if an unentangled state was prepared, the Hilbertspace
$\mathcal{H}_A$ would not be used at all and could as well be
omitted. In fact the four-dimensional space of two qubits turns
out to be necessary and sufficient to solve this case of the
problem.

If Alice makes a guess and prepares $\vert\psi\rangle_{AB}$ to be the
singlet state
\begin{equation}
\vert\psi\rangle_{AB}={1\over\sqrt{2}}(\ket{\uparrow\downarrow}_z-
    \ket{\downarrow\uparrow}_z) \label{singlet}
\end{equation}
which - up to a global phase factor - looks the same in all bases.
Here, the short hand notation $\ket{\uparrow \downarrow
}=\ket\uparrow\otimes\ket\downarrow$ is used. Then
$\vert\phi_{\eta_l}(\vec n_l)\rangle$  can be specified to be
proportional to $\lbrace{\ket{\uparrow\downarrow}}_x,
\vert\downarrow\uparrow\rangle_x,
\vert\uparrow\downarrow\rangle_y,
\vert\downarrow\uparrow\rangle_y,
\vert\uparrow\downarrow\rangle_z,\vert\downarrow\uparrow\rangle_z
\rbrace$. Now Alice wants to have the following look-up table:

\bigskip
\begin{center}
\mbox{
\begin{tabular}{|c||c|c|c|} \hline
         &     $x$  &    $y$  &    $z$     \\  \hline\hline
$\lambda_1$&$\;\;\;\;\downarrow\;\;\;\;$      & $\;\;\;\;\downarrow\;\;\;\;$&$\;\;\;\downarrow \;\;\;\;$\\  \hline
$\lambda_2$&$\uparrow$&$\uparrow$&$\downarrow$ \\  \hline
$\lambda_3$&$\downarrow$&$\uparrow$&$\uparrow$ \\  \hline
$\lambda_4$&$\uparrow$&$\downarrow$&$\uparrow$ \\  \hline
\end{tabular}}
\end{center}
\bigskip

That is, she wants
\begin{equation*}
     \vert\phi_1\rangle \;\; \bot \; \;\lbrace \ket{\uparrow,x},
     \ket{\uparrow,y},
     \ket{\uparrow,z} \rbrace,
\end{equation*}
\begin{equation*}
     \vert\phi_2\rangle \;\; \bot\;\; \lbrace \ket{\downarrow,x},
     \ket{\downarrow,y},
     \ket{\uparrow,z} \rbrace,
\end{equation*}
\begin{equation*}
     \vert\phi_3\rangle \;\;\bot \;\; \lbrace \ket{\uparrow,x},
     \ket{\downarrow,y}, \ket{\downarrow,z} \rbrace,
\end{equation*}
\begin{equation}
     \vert\phi_4\rangle \;\; \bot\;\; \lbrace \ket{\downarrow,x},
     \ket{\uparrow,y},
     \ket{\downarrow,z} \rbrace. \label{orthogonality conditions}
\end{equation}
and this choice of orthogonality
conditions (\ref{orthogonality conditions}) leads to the four orthonormal vectors
\begin{equation*}
     \vert\phi_1\rangle={1\over\sqrt{2}}\cdot\ket{\uparrow \downarrow}
     +{1\over 2}\cdot\left[\ket{\downarrow \downarrow}\cdot
     e^{i\pi\over4}-\ket{\uparrow \uparrow}\cdot e^{-i\pi\over4}\right]
\end{equation*}
\begin{equation*}
     \vert\phi_2\rangle={1\over\sqrt{2}}\cdot\ket{\uparrow
     \downarrow}-{1\over 2}\cdot
     \left[\ket{ \downarrow \downarrow}\cdot
     e^{i\pi\over4}-\ket{\uparrow \uparrow}\cdot e^{-i\pi\over4}\right]
\end{equation*}
\begin{equation*}
     \vert\phi_3\rangle={1\over\sqrt{2}}\cdot\ket{\downarrow\uparrow}
     +{1\over 2}\cdot\left[\ket{\downarrow \downarrow} \cdot
     e^{-i\pi\over4}-\ket{ \uparrow \uparrow}\cdot e^{i\pi\over4}\right]
\end{equation*}
\begin{equation}
     \vert\phi_4\rangle={1\over\sqrt{2}}\cdot\ket{\downarrow\uparrow
     }-{1\over 2}\cdot\left[\ket {\downarrow \downarrow }\cdot
     e^{-i\pi\over4}-\ket{\uparrow \uparrow }\cdot
     e^{i\pi\over4}\right] \label{seventeen}
\end{equation}
all expressed in the z-basis. (\ref{seventeen}) together with the state
(\ref{singlet}) and the above look-up table solves the problem.
(\ref{seventeen}) is very similar to the
result that is given in
\cite{Aharanov} and which is included in the appendix. In fact all possible
bases can be obtained from one by a unitary transformation, as we
shall see.

Now one could ask whether Alice really needs to have four
distinct eigenvalues $\lambda_j$ and therefore a big look-up
table, or whether the number of eigenvalues can be reduced.
However, the reduction of distinct eigenvalues turns out to be
impossible, whatever Hilbertspace $\mathcal{H}_{A}$ is used. To
prove this, we assume $\mathcal{H}_{A}$ to be $k$-dimensional with
basis  $\lbrace\vert i \rangle : 1 \leq i \leq k\rbrace$. Then
the state Alice prepares can be written as
\begin{eqnarray}
     \vert \psi\rangle_{AB}&=& \sum_{j=1}^k\sum_{\alpha\in\lbrace\uparrow,\downarrow\rbrace}a_{j,\alpha}\vert
     j
     \rangle\otimes\vert\alpha \rangle_z\nonumber\\
     &=&\sum_{j}a_{j,\uparrow}\vert j\rangle \otimes \vert \uparrow
     \rangle_z + \sum_{j} a_{j,\downarrow} \vert j
     \rangle \otimes \vert \downarrow \rangle_z\nonumber\\
     &=&{1\over\sqrt2}\cdot\left(\sum_{j}a_{j,\uparrow}\vert j
     \rangle \otimes \left(\vert \uparrow
     \rangle_x+\vert \downarrow
     \rangle_x\right) + \sum_{j} a_{j,\downarrow} \vert j
     \rangle \otimes  \left(\vert \uparrow
     \rangle_x-\vert \downarrow
     \rangle_x\right)\right)\nonumber\\
     &=&{1\over\sqrt2}\cdot\left(\sum_{j}a_{j,\uparrow}\vert j
     \rangle \otimes \left(\vert \uparrow\rangle_y+\vert \downarrow
     \rangle_y\right) + \sum_{j} a_{j,\downarrow} \vert j
     \rangle \otimes  \left(-i)(\vert \uparrow\rangle_y-\vert
     \downarrow\rangle_y\right)\right).
\end{eqnarray}
Bob's measurement leads to either of
\begin{eqnarray}
     \vert\phi_{+1}(\vec n_z)\rangle&=&\sum_{i}a_{i,\uparrow}\vert
     i\rangle \otimes \vert \uparrow\rangle_z\label{two}\\
     \vert\phi_{-1}(\vec
     n_z)\rangle&=&\sum_{i}a_{i,\downarrow}\vert i\rangle \otimes
     \vert \downarrow\rangle_z\label{three}\\
     \vert\phi_{+1}(\vec
     n_x)\rangle&=&\left(\sum_{i}a_{i,\uparrow}\vert i\rangle
     +\sum_{i}a_{i,\downarrow}\vert i
     \rangle\right)\otimes{1\over\sqrt2}\left(\vert \uparrow\rangle_z+\vert \downarrow
     \rangle_z\right)\nonumber\\
     &=&{1\over\sqrt2}\left(\sum_{i}a_{i,\uparrow}\vert i,
     \uparrow\rangle_z+\sum_{i}a_{i,\uparrow}\vert i,
     \downarrow\rangle_z
     +\sum_{i}a_{i,\downarrow}\vert i,
     \uparrow\rangle_z
     +\sum_{i}a_{i,\downarrow}\vert i,
     \downarrow\rangle_z\right)\label{four}\\
     \vert\phi_{-1}(\vec n_x)\rangle
     &=&\left(\sum_{i}a_{i,\uparrow}\vert i
     \rangle -\sum_{i}a_{i,\downarrow}\vert i\rangle\right)\otimes{1\over\sqrt2}(\vert \uparrow
     \rangle_z-\vert \downarrow\rangle_z)\nonumber\\
     &=&{1\over\sqrt2}\left(\sum_{i}a_{i,\uparrow}\vert i,
     \uparrow\rangle_z-\sum_{i}a_{i,\uparrow}\vert i,
     \downarrow\rangle_z
     -\sum_{i}a_{i,\downarrow}\vert i,
     \uparrow\rangle_z
     +\sum_{i}a_{i,\downarrow}\vert i,
     \downarrow\rangle_z\right)\label{five}\\
      \vert\phi_{+1}(\vec n_y)\rangle&=&\left(\sum_{i}a_{i,\uparrow}\vert i\rangle -i\sum_{i}a_{i,\downarrow}\vert i
      \rangle\right)\otimes{1\over\sqrt2}\left(\vert \uparrow\rangle_z+i\vert
     \downarrow\rangle_z\right)\nonumber\\
  &=&{1\over\sqrt2}\left(\sum_{i}a_{i,\uparrow}\vert i,
     \uparrow\rangle_z+i\sum_{i}a_{i,\uparrow}\vert i,
     \downarrow\rangle_z
     -i\sum_{i}a_{i,\downarrow}\vert i,
     \uparrow\rangle_z
     +\sum_{i}a_{i,\downarrow}\vert i,
     \downarrow\rangle_z\right)\label{six}\\
     \vert\phi_{-1}(\vec
     n_y)\rangle&=&\left(\sum_{i}a_{i,\uparrow}\vert i\rangle
     +i\sum_{i}a_{i,\downarrow}\vert i
     \rangle\right)\otimes{1\over \sqrt2}(\vert \uparrow\rangle_z-i\vert
     \downarrow\rangle_z)\nonumber\\
     &=&{1\over\sqrt2}\left(\sum_{i}a_{i,\uparrow}\vert i,
     \uparrow\rangle_z-i\sum_{i}a_{i,\uparrow}\vert i,
     \downarrow\rangle_z
     +i\sum_{i}a_{i,\downarrow}\vert i,
     \uparrow\rangle_z
     +\sum_{i}a_{i,\downarrow}\vert i,
     \downarrow\rangle_z\right)\label{seven}.
\end{eqnarray}
We see that (\ref{seven})=${\sqrt2}\cdot$(\ref{two})+${\sqrt2}\cdot$(\ref{three})-(\ref{six}) and (\ref{five})=${\sqrt2}\cdot$(\ref{two})+${\sqrt2}\cdot$(\ref{three})-(\ref{four}).
So we showed that the six results that can be obtained by Bob's
measurement are not linearly independent but lie in a
four-dimensional subspace of $\mathcal{H}_{AB}$. This however tells us
that the number of distinct  $\lambda_j$ has to be four. This can be
easily understood in the geometric picture adopted above. Assume there
are only three distinct $\lambda_j$. However big  $\mathcal{H}_{AB}$
is, the subspace that contains the states (\ref{two})-(\ref{seven}) is
four-dimensional and thus  $\mathcal{H}_{AB}$ can be assumed to be
four-dimensional as well. To ensure that Alice gets the look-up table she
wants, two of the four basis vectors $\ket{\phi_j}$ have to be
orthogonal to three states $\lbrace\vert\phi_{\eta_x}(\vec
     n_x)\rangle,\; \vert\phi_{\eta_y}(\vec
     n_y)\rangle,\; \vert\phi_{\eta_z}(\vec
     n_z)\rangle\rbrace$ with $\eta_l=1$ or $-1$. The latter however are linearly
independent and hence span a three-dimensional space. Therefore Alice
cannot find a two-dimensional subspace
orthogonal to all three.
This proves that Alice really needs four disinct values $\lambda_j$
and the size of the look-up table cannot be reduced.

We also note that as soon as $m\geq$3 we no longer have a mixture of
prediction and retrodiction. Alice can only retrodict the results Bob
obtained after having projected onto her final basis. If she
prepared $\vert\psi\rangle_{AB}$ to be an eigenstate of one of the
$\opone\otimes\hat{\vec\sigma}\cdot\vec n_l$ she would have to find out the result Bob
obtained by measuring along one of the two remaining axes by a single measurement,
which is impossible. On the other hand Alice cannot send an eigenstate of
two distinct spin operators even if entangled states are used, as we
have
$[\opone\otimes\hat\sigma_i,\opone\otimes\hat\sigma_j]=2i\varepsilon_{ijk}\opone\otimes\hat\sigma_k$,
and thus the measurement operators still do not commute.

\section{Non-orthogonal axes}

After having looked at the orthogonal case from as geometric
point of view, we will now tackle the case in which Bob can
measure along non-orthogonal axes and see that the number of axes
can even be increased. For that purpose we shall take a more
algebraic route, following \cite{Ben-Menahem}.

In the previous case in section 2, Alice's basis states
$\ket{\phi_j}$, and therefore the operator
$\widehat{\mathcal{M}}$, could be derived from the condition that
each of Alice's basis states $\ket{\phi_i}$ is orthogonal to
three of Bob's possible results $\vert\phi_{\eta_l}(\vec
n_l)\rangle$. Alice just guessed that $\ket\psi_{AB}$ might
either be (\ref{singlet}) or (\ref{fifteen}), hence calculated
$\vert\phi_{\eta_l}(\vec n_l)\rangle$ and via (\ref{orthogonality
conditions}) found a basis for $\widehat{\mathcal{M}}$. However,
to solve the problem in general, both $\ket\psi_{AB}$ and
$\widehat{\mathcal{M}}$ should be calculated, i.e.
$\ket\psi_{AB}$ should not be found by trial and error. To attack
this problem, we first note that Alice's initial state can be
written as
\begin{equation}
     \ket\psi_{AB}=\sum_{j=1}^{2k} b_j\ket{\phi_j},\label{phiAB}
\end{equation}
where $\{\ket{\phi_j}\}$ is a basis of $\mathcal{H}_{AB}$. Without
loss of generality the $b_j$ can be assumed to be real. Then the
problem reduces to finding the basis states $\ket{\phi_j}$
together with the values $b_j$, as these give both
$\widehat{\mathcal{M}}$ and $\ket\psi_{AB}$. To find these we
look at the conditions they have to satisfy, to ensure Alice can
retrodict Bob's results. First of all $ \ket\psi_{AB}$ should be
normalised, which gives:
\begin{equation}
      \sum_jb_j^2=1 \label{eleven}.
\end{equation}
A second condition is that, in order to enable Alice to retrodict
Bob's results $\eta_l$ deterministically, the states
$\ket{\phi_{+1}(\vec n_l)}$ and $\ket{\phi_{-1}(\vec n_l)}$ must
lie in the span of disjoint subsets of
$\left\lbrace\ket{\phi_j}\right\rbrace$. Otherwise the result
$\lambda_i$, which Alice gets if she measures using $\mathcal{M}$,
would not tell her $\eta_l$. As this must be true for all axes
$l$ along which Bob can measure, there are $m$ pairs of disjoint
subsets. Therefore, let us introduce $2m$ sets of indices
$S_{\eta_l}(\vec n_l)$ that indicate which basis vectors span the
vectors
\begin{equation}
     \ket{\phi_{\eta_l}(\vec n_l)}=\sum_{j\in
     S_{\eta_l}(\vec n_l)} b_j\ket{\phi_j}. \label{phietal}
\end{equation}
Here the $b_j$ are the same as in (\ref{phiAB}), which can be seen from
\begin{equation}
      \vert\psi\rangle_{AB}=  \vert\phi_+(\vec n_l)\rangle+ \vert\phi_-(\vec n_l)\rangle. \label{sumeta}
\end{equation}
This equation only states that Bob always gets either `up' or
`down'. So the $S_{\eta_l}(\vec n_l)$ provide a look-up table for
Alice. In the case she measures $\lambda_i$ corresponding to
$\ket{\phi_i}$, Alice checks whether $i\in S_{+1}(\vec n_l)$ or $i
\in S_{-1}(\vec n_l)$ which tells her which result Bob got if he
measured along $\vec n_l$.

For further convenience it is helpful to introduce a sign function for the partitions  $S_{\eta_l}(\vec n_l)$ in the following way:
\begin{equation}
    \varepsilon_j^{(l)}= \left\lbrace
        \begin{matrix}
               +1\;\;\;\mbox{for}\;\; j\in S_{+1}(\vec n_l)\\
               -1\;\;\; \mbox{for}\;\; j\in S_{-1}(\vec n_l)
        \end{matrix}\right.
\end{equation}
For example,  if $\ket{\phi_{+1}(\vec
n_l)}=b_1\ket{\phi_1}+b_3\ket{\phi_3}+b_6\ket{\phi_6}$,
$\varepsilon_1^{(l)}$, $\varepsilon_3^{(l)}$ and
$\varepsilon_6^{(l)}$ give +1, the others give -1.

Now some algebraic manipulations have to be done to obtain
sufficient information for Alice to construct her states. The
basic task is to find some constraints for the look-up table that
Alice chooses. Once a ``good'' table is chosen, the basis states
can be constructed. From (\ref{sumeta}) we get (recalling
(\ref{one})):
\begin{equation}
     \opone\otimes(\hat\sigma\cdot\vec
     n_l)\vert\psi\rangle_{AB}=\sum_j\epsilon_j^{(l)}b_j\vert\phi_j\rangle\;\;\;\;\;\;\;
     \mbox{for} \;\;\; 1 \leq l \leq m \label{thirteen}.
\end{equation}
Furthermore the Pauli-matrices satisfy\footnote{Note that $(\widehat {\opone\otimes\vec\sigma}) \cdot \vec
     n_a=\opone\otimes (\hat {\vec\sigma} \cdot \vec n_a)$.}:
\begin{equation}
     ((\widehat {\opone\otimes\vec\sigma}) \cdot \vec
     n_a)\cdot((\widehat{\opone\otimes\vec\sigma}) \cdot \vec n_b)=
     \vec n_a\cdot\vec n_b+i(\vec n_a\times \vec n_b)\cdot \widehat{\opone\otimes\vec\sigma}.\label{eight}
\end{equation}
The expectation value of (\ref{eight}) yields:
\begin{equation}
     _{AB}\langle\psi\vert \vec n_a\cdot\vec n_b\vert\psi\rangle_{AB}+i_{AB}\langle\psi\vert(\vec n_a\times \vec n_b)\cdot \widehat{\opone\otimes\vec\sigma}  \vert\psi\rangle_{AB}=\sum_j\varepsilon_j^{(a)}\varepsilon_j^{(b)}b_j^2
\end{equation}
and therefore the two equations,
\begin{equation}
    0=_{AB}\langle\psi\vert   (\vec n_l \cdot(\widehat
     {\opone\otimes\vec\sigma}))\vert\psi\rangle_{AB}= \sum_j\varepsilon_j^{(l)}b_j \label{nine}
\end{equation}
and
\begin{equation}
     \vec n_l\cdot \vec n_j=\sum_s\varepsilon_s^{(l)}\varepsilon_s^{(j)}b_s^2\label{ten},
\end{equation}
for the real and imaginary components. From (\ref{nine}) we obtain:
\begin{equation}
\sum_{j \in S_{\eta}}b_j^2={1\over2} \label{fourteen},
\end{equation}
which tells us that the square of the coefficients of the basis states $\ket{\phi_j}$ have to add to ${1\over2}$
for each disjoint subset.
Finally we note that in the case in which $m>3$, of course three axes are sufficient to span all the others:
\begin{equation}
     \vec n_{k+3}=\sum_{l=1}^3c_l^{(k)}\vec n_l \;\;\;\;\;\; \mbox{for}
     \;\;\;1 \leq k \leq m-3,
\end{equation}
and thus from (\ref{thirteen}):
\begin{equation}
     \varepsilon_s^{(k+3)}=\sum_{l=1}^3c_l^{(k)}\varepsilon_s^{(l)}\;\;\;\;\; \mbox{for}\;\;\;\;\;1\leq s \leq 2k. \label{twelve}
\end{equation}

It turns out that these relations enable Alice to calculate both
the $b_j$ and the $\ket{\phi_j}$. It was shown in
\cite{Ben-Menahem} that as a necessary and sufficient condition
for the construction of a basis we need (i) $\varepsilon_l^{(l)}$
and $b_j$ that solve (\ref{eleven}), (\ref{nine}) and
(\ref{ten}); and (ii) if $m\geq$3, there exist numbers $c_l^{(k)}$
$l \leq l\leq 3$ and $l\leq k\leq m-3$ such that (\ref{twelve})
holds. Furthermore it was proved that no solutions exist for
$m>4$ and that if $m=4$ the directions have to satisfy
$\sum_{l=1}^4 \vec n_l=0$. But all this only means that the
look-up table has to satisfy certain conditions. This, however,
was anticipated already as not any arbitrary look-up table should
lead to a solution.

However, once Alice chooses a suitable look-up table, the basis
states can be derived. The orientation of Bob's axes (\ref{ten}),
together with Alice condition (\ref{fourteen}), lead to a unique
set of components $b_j$. The crucial equation (\ref{thirteen})
now tells us the action of Bob`s measurement on $\ket\psi_{AB}$.
These can be inverted to obtain $\lbrace\ket{\phi_j}\rbrace$. The
details of this process can best be understood by looking at an
example.

In the following we will explicitely construct the basis for the case m=4, using
this method.
The non-orthogonal $m=3$ case can be solved equivalently and a possible
basis is given in the appendix.

The problem is solved using a six-dimensional Hilbert space $\mathcal{H}_{AB}$. A basis of this space is denoted by
$\lbrace\ket{2,1},\ket{2,-1},...,\ket{0,-1}$ where $\ket{\rho,\zeta}=\ket{\rho}\otimes\ket{\zeta}$ with $\rho =
1,2,3$ and $\zeta=\pm1$ is the tensor product of a \textit{qutrit} and a qubit.

We use the partition provided in \cite{Ben-Menahem}, namely
\begin{equation*}
      S_+(\vec n_1) = \lbrace 1,2,3 \rbrace,
\end{equation*}
\begin{equation*}
      S_+(\vec n_2)=\lbrace 1,5,6 \rbrace,
\end{equation*}
\begin{equation}
      S_+(\vec n_3)=\lbrace 3,4,6 \rbrace.
\end{equation}
which satisfy the necessary conditions. From (\ref{twelve}) we get
$\varepsilon_1^{(4)}=\varepsilon_3^{(4)}=\varepsilon_6^{(4)}=-1;\;\varepsilon_2^{(4)}=\varepsilon_4^{(4)}=\varepsilon_5^{(4)}=1$.
This is nothing but the following look-up table for Alice:

\bigskip
\begin{center}
\mbox{
\begin{tabular}{|c||c|c|c|c|} \hline
         &      $\vec n_1$   &    $\vec n_2$  &     $\vec n_3$  &    $\vec n_4$   \\           \hline \hline
$\lambda_1$ & $\;\;\;\;\uparrow\;\;\;\;$ & $\;\;\;\;\uparrow\;\;\;\;$ & $\;\;\;\;\downarrow\;\;\;\;$ & $\;\;\;\;\downarrow\;\;\;\;$ \\  \hline
$\lambda_2$ & $\uparrow$ & $\downarrow$ & $\downarrow$ & $\uparrow$   \\  \hline
$\lambda_3$ & $\uparrow$ & $\downarrow$ & $\uparrow$ & $\downarrow$   \\  \hline
$\lambda_4$ & $\downarrow$ & $\downarrow$ & $\uparrow$ & $\uparrow$   \\  \hline
$\lambda_5$&$\downarrow$&$\uparrow$&$\downarrow$&$\uparrow$ \\    \hline
$\lambda_6$&$\downarrow$&$\uparrow$& $\uparrow$  & $\downarrow$     \\ \hline
\end{tabular}}
\end{center}

\bigskip
From (\ref{fourteen}) this table gives immediatly
\begin{equation}
      b_1^2+ b_2^2+ b_3^2= b_4^2+ b_5^2+ b_6^2= b_1^2+ b_5^2+ b_6^2=
      b_3^2+ b_4^2+ b_6^2={1\over 2}.
\end{equation}
These equations can be solved in terms of the two free parameters $b_5$ and $b_6$:
\begin{equation}
      b_1^2= b_4^2={1\over2}-b_5^2-b_6^2;\;\;\;\; b_2^2= b_6^2;\;\;\;\; b_3^2=b_5^2.
\end{equation}
(\ref{ten}) yields:
\begin{equation}
      \vec n_1\cdot\vec n_2=1-4b_5^2-4b_6^2;\;\;\;\;
      \vec n_2\cdot\vec n_3=4b_6^2-1;\;\;\;\;
      \vec n_3\cdot\vec n_1=4b_5^2-1.
\end{equation}
Therefore the two parameters $b_5,\;b_6$ specify the mutual orientation of the axes.
These are all the preliminaries we need to construct our basis.
To specify the axes we make the symmetric choice:
\begin{equation*}
      b_j={1\over\sqrt6}\;\;\;\;\; \forall j.
\end{equation*}
The crucial step is that (\ref{thirteen}) tells us the action of $\hat\sigma_x,\hat\sigma_y,\hat\sigma_z$ on
$\vert \psi\rangle_{AB}$, giving four orthonormal states:
\begin{eqnarray}
  \vert\psi\rangle_{AB}&=&{1\over\sqrt6}(\vert\phi_1\rangle+\vert\phi_2\rangle+\vert\phi_3\rangle+\vert\phi_4\rangle+\vert\phi_5\rangle+\vert\phi_6\rangle)\\
(\opone\otimes\hat\sigma_x)
  \vert\psi\rangle_{AB}&=&{1\over\sqrt6}(\vert\phi_1\rangle+\vert\phi_2\rangle+\vert\phi_3\rangle-\vert\phi_4\rangle-\vert\phi_5\rangle-\vert\phi_6\rangle) \\
  (\opone\otimes
  \hat\sigma_y)\vert\psi\rangle_{AB}&=&{1\over\sqrt{12}}(2\cdot\vert\phi_1\rangle-\vert\phi_2\rangle-\vert\phi_3\rangle-2\cdot\vert\phi_4\rangle+\vert\phi_5\rangle+\vert\phi_6\rangle) \\
 (\opone\otimes\hat\sigma_z)\vert\psi\rangle_{AB}&=&{1\over
    2}(-\vert\phi_2\rangle+\vert\phi_3\rangle-\vert\phi_5\rangle+\vert\phi_6\rangle)
\end{eqnarray}
These are orthonormal and can be extended to a basis by
\begin{equation} \vert\chi_1\rangle={1\over
    2}\cdot(-\vert\phi_1\rangle+\vert\phi_2\rangle-\vert\phi_4\rangle+\vert\phi_6\rangle)
\end{equation}
and
\begin{equation} \vert\chi_2\rangle={1\over
    12}\cdot(-\vert\phi_1\rangle-\vert\phi_2\rangle + 2 \cdot\vert\phi_3\rangle-\vert\phi_4\rangle
   + 2\cdot\vert\phi_5\rangle-\vert\phi_6\rangle).
\end{equation}
These equations have to be inverted to get explicit expressions
for the basis states $\ket{\phi_j}$. In order to express them in
terms of the basis $\lbrace\ket{2,1},\ket{2,-1},...,\ket{0,-1}$
of the six-dimensional vector space $\mathcal{H}_{AB}$, we note
that ${1\over \sqrt2}\cdot((\opone\otimes\hat\sigma_x)+i\zeta
(\opone\otimes\hat\sigma_y))\ket{\psi}_{AB}$ is an eigenstate of
$(\opone\otimes\hat\sigma_z)$ with eigenvalue $\zeta=\pm1$, as
well as ${1\over \sqrt2}\cdot((\opone\otimes\opone+\zeta
(\opone\otimes\hat\sigma_y))\ket{\psi}_{AB}$ and
($\vert\chi_1\rangle\cdot\delta_{1\zeta}+\vert\chi_2\rangle\cdot\delta_{-1\zeta}$),
with the usual Kronecker symbol $\delta_{ij}$. Hence these states
form a basis of the three-dimensional eigenspace of the operator
$\opone\otimes\hat\sigma_z$ with eigenvalue $\zeta$. However
{$\vert 2,\zeta\rangle,\vert 1,\zeta\rangle,\vert
0,\zeta\rangle,$} is also a basis of this subspace, therefore
there has to be a unitary transformation, such that
\begin{equation}
     \left(
           \begin{matrix}
                {1\over \sqrt2}\cdot((\opone\otimes\hat\sigma_x)+i\zeta(\opone\otimes\hat\sigma_y))\ket{\psi}_{AB}\\
                {1\over \sqrt2}\cdot(\opone\otimes\opone+\zeta(\opone\otimes\hat\sigma_y))\ket{\psi}_{AB}\\
                \vert\chi_1\rangle\cdot\delta_{1,\zeta}+\vert\chi_2\rangle\cdot\delta_{-1,\zeta}
           \end{matrix}
     \right)= e^{i \hat \theta_{\zeta} \cdot \hat \tau + i\Lambda_{\zeta}}\cdot
     \left(
           \begin{matrix}
                \vert 2,\zeta\rangle\\
                \vert 1,\zeta\rangle\\
                \vert0,\zeta\rangle
           \end{matrix}
     \right). \label{unitary}
\end{equation}
Here $\hat \theta_{\pm}$ are arbitrary eight-vectors, and $\hat
\tau$ is a vector consisting of the eight generators of SU(3);
$\Lambda_{\pm}$ are arbitrary numbers. Thus, the most general
solution is characterised by a set of 24 numbers $( \vec
\theta_{\pm},\Lambda_{\pm},\lbrace\lambda_j\rbrace).$ Equation
(\ref{unitary}) allows us to express the basis states
$\ket{\phi_j}$ in terms of the computational basis of
$\mathcal{H}_{AB}$. Without loss of generality, we choose
$\vec\theta_{\zeta}$=0 and $\Lambda_{\zeta}$=0 for both $\zeta$
and finally get:

\begin{eqnarray*}
     \vert\phi_1\rangle&=& \left({1\over\sqrt12}-{i\over\sqrt6}\right)\vert
     2,\uparrow\rangle + {1\over\sqrt12}\vert
     1,\uparrow\rangle - {1\over2} \vert 0,\uparrow \rangle+\left({1\over\sqrt12}+{i
     \over\sqrt6}\right)\vert 2,\downarrow \rangle+ {1\over\sqrt12}\vert
     1,\downarrow\rangle- {1\over\sqrt12} \vert 0, \downarrow \rangle\\
     \vert\phi_2\rangle& =& \left({-1\over\sqrt12}+{i\over\sqrt6}\right)\vert
     2,\uparrow\rangle + {1\over\sqrt12}\vert
     1,\uparrow\rangle - {1\over2} \vert 0,\uparrow \rangle+\left({-1\over\sqrt12}-{i
     \over\sqrt6}\right)\vert 2,\downarrow \rangle+ {1\over\sqrt12}\vert
     1,\downarrow\rangle- {1\over\sqrt12} \vert 0, \downarrow
     \rangle\\
     \vert\phi_3\rangle& =& \left({1\over\sqrt12}+{i\over\sqrt24}\right)\vert
     2,\uparrow\rangle + \left({1\over\sqrt12}+{1\over\sqrt8}\right)\vert
     1,\uparrow\rangle +\left({1\over\sqrt12}-{i
     \over\sqrt24}\right)\vert 2,\downarrow \rangle+ \left({1\over\sqrt12}-{1\over\sqrt8}\right)\vert
     1,\downarrow\rangle+ {1\over\sqrt3} \vert 0, \downarrow \rangle\\
     \vert\phi_4\rangle&=& \left({-1\over\sqrt12}-{i\over\sqrt24}\right)\vert
     2,\uparrow\rangle + \left({1\over\sqrt12}-{1\over\sqrt8}\right)\vert
     1,\uparrow\rangle +\left({-1\over\sqrt12}+{i
     \over\sqrt24}\right)\vert 2,\downarrow \rangle+ \left({1\over\sqrt12}+{1\over\sqrt8}\right)\vert
     1,\downarrow\rangle+ {1\over\sqrt3} \vert 0, \downarrow \rangle\\
     \vert\phi_5\rangle&=& \left({-1\over\sqrt12}-{i\over\sqrt24}\right)\vert
     2,\uparrow\rangle + \left({1\over\sqrt12}+{1\over\sqrt8}\right)\vert
     1,\uparrow\rangle + {1\over2} \vert 0,\uparrow \rangle+\left({-1\over\sqrt12}+{i
     \over\sqrt24}\right)\vert 2,\downarrow \rangle+ \left({1\over\sqrt12}-{1\over\sqrt8}\right)\vert
     1,\downarrow\rangle- {1\over\sqrt12} \vert 0, \downarrow \rangle\\
     \vert\phi_6\rangle& =& \left({1\over\sqrt12}+{i\over\sqrt24}\right)\vert
     2,\uparrow\rangle + \left({1\over\sqrt12}+{1\over\sqrt8}\right)\vert
     1,\uparrow\rangle + {1\over2} \vert 0,\uparrow \rangle+\left({1\over\sqrt12}-{i
     \over\sqrt24}\right)\vert 2,\downarrow \rangle+ \left({1\over\sqrt8}+{1\over\sqrt12}\right)\vert
     1,\downarrow\rangle- {1\over\sqrt12} \vert 0, \downarrow \rangle,
\end{eqnarray*}
all expressed in the z-basis.

So we have explicitly constructed a basis for a special symmetric
orientation of our axes. This technique can be applied to all
possible orientations that satisfy the condition $\sum_{l=1}^{4}
\vec n_l=0$.

\section{The Quantum Network}

After having explicitly constructed some measurement bases, we now
want to know how Alice's measurements can actually be performed.
Thus, we want to construct a device that prepares Alice's initial
state, and the operator $\widehat{\mathcal{M}}$ that projects onto
one of the bases that were just determined.

In principle, this can be achieved by constructing a quantum
network which is a kind of building plan for Alice's devices. For
that purpose we regard the Hilbert space $\mathcal{H}_{AB}$ as a
subspace of the $2^N$-dimensional space
$\mathcal{H}=\mathcal{H}_B\otimes\ \mathcal{H}_B\otimes...\otimes
\mathcal{H}_B$. The basis of this space is taken to be the tensor
product of the qubit basis of $\mathcal{H}_B$, i.e. the states
$\vert\uparrow\uparrow...\uparrow\rangle,\;
\vert\uparrow\uparrow...\downarrow\rangle$,\; ... \footnote{It is
convenient to use $\ket{0}:=\ket{\uparrow}$ and
$\ket{1}:=\ket{\downarrow}$ in this section.}. In this space of
qubits we can now perform certain operations and in that way try
to construct the whole experimental device out of a few basic
transformations which we call quantum gates. As we demand
reversibility, quantum gates have to be unitary. (We could also
say, that as we have \textit{quantum} gates which obey the rules
of quantum mechanics, we have unitary transformations and
therefore reversibility). It was shown in \cite{IBM} that there
is a set of universal quantum gates from which every evolution of
a quantum system can be approximated with arbitrary precision. A
set of universal gates can be taken to be the \textit{Phase
Shift} $P$($\phi$), defined by
\begin{eqnarray}
      \vert 0\rangle& \mapsto& \vert 0\rangle\nonumber\\
      \vert 1\rangle& \mapsto& e^{i\phi}\vert 1\rangle,
\end{eqnarray}
the \textit{Controlled Not} ($CNOT$),
\begin{eqnarray}
      \vert 00\rangle& \mapsto& \vert 00\rangle\nonumber\\
      \vert 01\rangle& \mapsto& \vert 01\rangle\nonumber\\
      \vert 10\rangle& \mapsto& \vert 11\rangle\nonumber\\
      \vert 11\rangle& \mapsto& \vert 10\rangle,
\end{eqnarray}
and the \textit{Hadamard gate} ($H$),
\begin{eqnarray}
      \vert 0\rangle& \mapsto& {1\over\sqrt2}(\vert0\rangle+\vert 1\rangle)\nonumber\\
      \vert 1\rangle& \mapsto& {1\over\sqrt2}(\vert0\rangle-\vert 1\rangle).
\end{eqnarray}
If we use the representation,
\begin{equation}
     \ket0=\left(
     \begin{array}{cc}
        1 \\
        0
     \end{array}
     \right),\;
     \ket1=\left(
     \begin{array}{cc}
        0 \\
        1
     \end{array}
     \right),\;
     \ket{00}=\ket0\otimes \ket0=\left(
     \begin{array}{c}
        1\\
        0\\
        0\\
        0
     \end{array}
     \right),...,
\end{equation}
the basic transformations can be written as:
\begin{equation}
     P(\phi)=\left(
     \begin{array}{cc}
        1 & 0 \\
        0 & e^{i\phi}
     \end{array}
     \right),\;
     H=\frac{1}{\sqrt{2}}\left(
     \begin{array}{cc}
        1 & 1 \\
        1 & -1
     \end{array}
     \right),\;
     CNOT=\left(
     \begin{array}{cccc}
        1 & 0 & 0 & 0\\
        0 & 1 & 0 & 0\\
        0 & 0 & 0 & 1\\
        0 & 0 & 1 & 0
\end{array}
     \right).
\end{equation}
Note that these gates only act on one or two qubits. The fact that
they are nevertheless universal is very important as these gates
can, in principle, be built and therefore even complicated
calculations can be done without using gates that act on many
qubits at once.

These gates can also be visualized in an elegant way. We denote
the \textit{Phase Shift} by

\setlength{\unitlength}{0.030in}
\begin{equation}
\mbox{\hspace{3cm}}
\mbox{
\begin{picture}(30,0)(15,15)
  \put(-4,14){$\ket{x}$} \put(5,15){\line(1,0){20}} \put(20,15){\line(1,0){5}}
  \put(15,15){\circle*{3}}
\put(14,19){$\phi $}
\put(30,14){$e^{ix\phi}\ket{x}$}
\end{picture}
}, \label{Phase Shift}
\end{equation}

\bigskip
the \textit{Hadamard gate} by
\setlength{\unitlength}{0.030in}
\begin{equation}
 \mbox{\hspace{3cm}} \mbox{
\begin{picture}(60,0)(15,15)
  \put(-4,14){$\ket{x}$} \put(5,15){\line(1,0){5}} \put(20,15){\line(1,0){5}}
  \put(10,10){\framebox(10,10){$H$}}
\put(30,14){$\displaystyle\frac{(-1)^x\ket{x}+\ket{1-x}}{\sqrt{2}}$}
\end{picture}
} \label{Hadamard Gate}
\end{equation}

\bigskip
and the \textit{Controlled Not} by

\begin{equation}
\mbox{\hspace{4cm}}
\mbox{
\begin{picture}(30,10)(20,15)
  \put(-4,14){$\ket{y}$} \put(-4,29){$\ket{x}$} \put(5,15){\line(1,0){20}}
\put(5,30){\line(1,0){20}}
  \put(15,30){\circle*{3}} \put(15,15){\line(0,1){15}}
\put(15,15){\circle{3}} \put(30,21){ $|\,x\rangle |\,y+x\rangle$,}
\end{picture}}\label{Controlled Not}
\end{equation}
where on the left hand side of the diagram the input state is
given, the output on the right, and $x,y\in\{0,1\}$. Finally we
note that the \textit{Controlled Phase Shift CP($\phi$)} can be
built out of the above states. It is denoted by

\begin{eqnarray}
{CP}(\phi )= \left(
\begin{array}{cccc}
1 & 0 & 0 & 0 \\
0 & 1 & 0 & 0 \\
0 & 0 & 1 & 0 \\
0 & 0 & 0 & e^{i\phi }
\end{array}
\right)
\mbox{\hspace{1.5cm}}
\mbox{
\begin{picture}(25,0)(0,20)
  \put(-4,14){$\ket{y}$} \put(-4,29){$\ket{x}$} \put(5,15){\line(1,0){20}}
\put(5,30){\line(1,0){20}}
  \put(15,30){\circle*{3}} \put(15,15){\line(0,1){15}} \put(17,21){$\phi$}
\put(15,15){\circle*{3}}
\end{picture}
}e^{ixy\phi} | \,x\rangle |\,y \rangle . \label{Controlled Phase Shift}
\end{eqnarray}

After having set up the basic framework we will present networks
that prepare Alice's initial state and project onto the basis
$\ket{\phi_j}$. Let us first look at the case in which the
symmetric state (\ref{fifteen}) is prepared, which is projected
onto (\ref{sixteen}). In this case we have the following network:

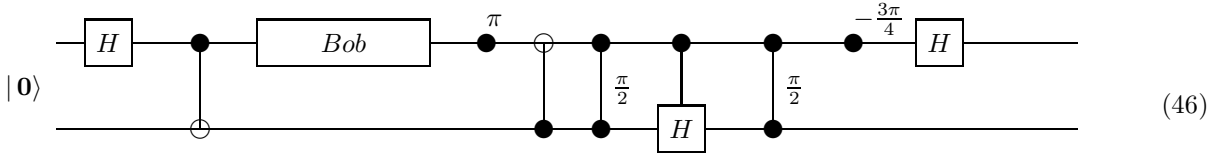
\begin{equation}
\mbox{\hspace{5cm}}
\mbox{
\begin{picture}(300,0)(50,25)
\put(-4,21){$\ket{\bold 0}$}
\put(5,15){\line(1,0){105}}
\put(5,30){\line(1,0){5}}
\put(18,30){\line(1,0){22}}
\put(70,30){\line(1,0){85}}
\put(163,30){\line(1,0){20}}
\put(10,26){\framebox(8,8){$H$}}
  \put(30,30){\circle*{3}} \put(30,15){\line(0,1){15}}
\put(30,15){\circle{3}}
\put(40,26){\framebox(30,8){$Bob$}}
\put(80,30){\circle*{3}} \put(80,33){$\pi$}
\put(90,15){\line(0,1){15}}
\put(90,30){\circle{3}}
\put(90,15){\circle*{3}}
\put(100,30){\circle*{3}} \put(100,15){\line(0,1){15}}
\put(100,15){\circle*{3}}
\put(130,30){\circle*{3}} \put(130,15){\line(0,1){15}}
\put(130,15){\circle*{3}}
\put(144,30){\circle*{3}}
\put(114,30){\circle*{3}}
\put(110,11){\framebox(8,8){$H$}}
\put(155,26){\framebox(8,8){$H$}}
\put(118,15){\line(1,0){32}}
\put(114,19){\line(0,1){11}}
\put(150,15){\line(1,0){33}}
\put(102,21){$\pi\over2$}
\put(132,21){$\pi\over2$}
\put(144,33){$-{3\pi\over4}$}
\end{picture}}\label{qgate1}
\end{equation}

\bigskip
\bigskip
We use the convention that the original state is described by
$\vert\bold 0\rangle=\vert 0\rangle\otimes\vert 0\rangle\otimes...\otimes\vert 0\rangle\in (\mathcal{H}_B)^N=\mathcal{H}$.
Then the first part of the network gives:
\begin{equation}
      \vert \bold 0\rangle=\vert0\rangle\otimes\vert0\rangle\;\;\stackrel{H_1}{\mapsto}\;\;{1\over\sqrt2}(\vert00\rangle+\vert10\rangle)\;\;\stackrel{CN_2}{\mapsto}
      \;\;{1\over\sqrt2}(\vert00\rangle+\vert11\rangle).
\end{equation}
Here the indices denote the qubit on which the transformation is
performed. To prove that the second part projects onto the basis
(\ref{sixteen}), we have to show that the set of basis vectors in
(\ref{sixteen}) is mapped bijectively onto
$\lbrace\vert00\rangle,\vert01\rangle,\vert10\rangle,\vert11\rangle
\rbrace$. For that purpose we introduce yet another gate, the
\textit{Controlled Unitary gate ($CU$)}:

\begin{equation}
\mbox{\hspace{5cm}}
\mbox{
\begin{picture}(70,0)(50,25)
\put(5,33){\line(1,0){18}}
\put(5,18){\line(1,0){5}}
\put(18,18){\line(1,0){5}}
\put(10,14){\framebox(8,8){$U$}}
\put(14,33){\circle*{3}} \put(14,22){\line(0,1){11}}
\put(30,25){=}
\put(45,33){\circle*{3}} \put(45,18){\line(0,1){15}}
\put(45,18){\circle*{3}}
\put(75,33){\circle*{3}} \put(75,18){\line(0,1){15}}
\put(75,18){\circle*{3}}
\put(84,33){\circle*{3}}
\put(60,33){\circle*{3}}
\put(56,14){\framebox(8,8){$H$}}
\put(64,18){\line(1,0){26}}
\put(60,22){\line(0,1){11}}
\put(47,24){$\pi\over2$}
\put(77,24){$\pi\over2$}
\put(84,37){$-{3\pi\over4}$}
\put(40,33){\line(1,0){49}}
\put(40,18){\line(1,0){16}}
\end{picture}}
CU=\left(
\begin{array}{cccc}
1 & 0 & 0 & 0 \\
0 & 1 & 0 & 0 \\
0 & 0 & {e^{-i{3\pi\over4}}\over\sqrt2} & i\cdot {e^{-i{3\pi\over4}}\over\sqrt2} \\
0 & 0 & i\cdot {e^{-i{3\pi\over4}}\over\sqrt2} & {e^{-i{3\pi\over4}}\over\sqrt2}
\end{array}
\right).
\end{equation}

The \textit{Contolled Hadamard} ($CH$) can be built directly.

Now we can easily check:
\begin{eqnarray}
    \vert\phi_{1,2}\rangle&=&{1\over\sqrt{2}}\vert00
    \rangle\pm{1\over 2}\left[\vert01\rangle\cdot
    e^{i\pi\over4}+\vert10\rangle\cdot
    e^{-i\pi\over4}\right]\nonumber\\
    &\stackrel{P(\pi)_1}{\mapsto}&{1\over\sqrt{2}}\vert00
    \rangle\pm{1\over 2}\left[\vert01\rangle\cdot
    e^{i\pi\over4}-\vert10\rangle\cdot
    e^{-i\pi\over4}\right] \nonumber\\
    &\stackrel{CN_1}{\mapsto}&{1\over\sqrt2}\vert00\rangle\pm{1\over2}\left[\vert11\rangle\cdot
    e^{i\pi\over4}-\vert10\rangle\cdot e^{-i\pi\over4}\right]\nonumber\\
    &\stackrel{CU_2}{\mapsto}&  {1\over\sqrt2}\vert00\rangle\pm{1\over2}\left[{1\over\sqrt2}(\vert11\rangle+i\vert10\rangle)\cdot
    e^{-i\pi\over2}-{1\over\sqrt2}(\vert10\rangle+i\vert11\rangle)\cdot e^{-i\pi}\right]={1\over\sqrt2}\vert00\rangle\pm
    \vert10\rangle)\nonumber\\
    &\stackrel{H_1}{\mapsto}&\left\{
    \begin{matrix}
            \vert00\rangle\;\;\;\;\mbox{for}\;\;\;\;  \vert\phi_{1}\rangle\\
            \vert10\rangle\;\;\;\;\mbox{for} \;\;\;\; \vert\phi_{2}\rangle
    \end{matrix}\right.\nonumber\\
    \vert\phi_{3,4}\rangle&=&{1\over\sqrt{2}}\ket{11}\pm{1\over 2}\left[\ket{01}\cdot
    e^{i\pi\over4}+\ket{10}\cdot
    e^{-i\pi\over4}\right] \nonumber\\
    &\stackrel{P(\pi)}{\mapsto}&{1\over\sqrt2}\vert11\rangle\pm{1\over2}\left[\vert01\rangle\cdot
    e^{i\pi\over4}-\vert10\rangle\cdot e^{-i\pi\over4}\right] \nonumber\\
    &\stackrel{CN_1}{\mapsto}&{1\over\sqrt2}\vert01\rangle\pm{1\over2}\left[\vert11\rangle\cdot
    e^{i\pi\over4}-\vert10\rangle\cdot e^{-i\pi\over4}\right]\nonumber\\
    &\stackrel{CU_2}{\mapsto}&
    {1\over\sqrt2}\vert01\rangle\pm{1\over2}\left[{1\over\sqrt2}(\vert11\rangle+i\vert10\rangle)\cdot e^{-i{\pi
    \over 2}}-{1\over\sqrt(2)}(\vert10\rangle+i\vert11\rangle)\cdot e^{-i\pi}\right]={1\over\sqrt2}\vert00\rangle\pm \vert10\rangle)\nonumber\\
    &\stackrel{H_1}{\mapsto}&\left\{
    \begin{matrix}
            \vert01\rangle\;\;\;\;\mbox{for}\;\;\;\;  \vert\phi_{3}\rangle\\
            \vert11\rangle\;\;\;\;\mbox{for} \;\;\;\; \vert\phi_{4}\rangle
    \end{matrix}\right.\nonumber
\end{eqnarray}

If Alice prepares the singlet state (\ref{singlet}), the gate looks slightly
more complicated:

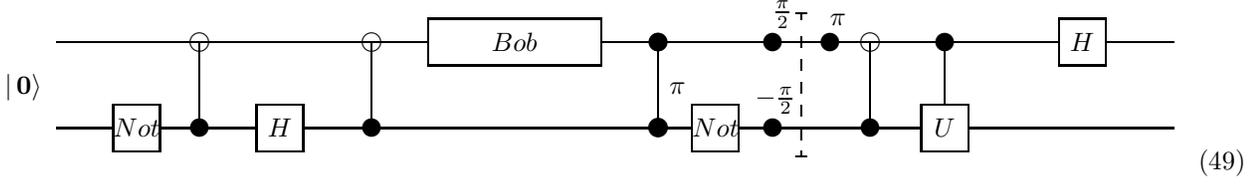
\begin{equation}
\mbox{\hspace{4,5cm}}
\mbox{
\begin{picture}(300,20)(50,15)
\put(-4,21){$\ket{\bold 0}$} \put(5,15){\line(1,0){10}}
\put(23,15){\line(1,0){17}} \put(48,15){\line(1,0){68}}
\put(5,30){\line(1,0){65}} \put(15,11){\framebox(8,8){$Not$}}
\put(48,30){\line(1,0){22}} \put(100,30){\line(1,0){80}}
\put(188,30){\line(1,0){12}} \put(40,11){\framebox(8,8){$H$}}
\put(30,30){\circle{3}} \put(30,15){\line(0,1){15}}
\put(30,15){\circle*{3}} \put(60,30){\circle{3}}
\put(60,15){\line(0,1){15}} \put(60,15){\circle*{3}}
\put(70,26){\framebox(30,8){$Bob$}} \put(110,30){\circle*{3}}
\put(110,15){\line(0,1){15}} \put(110,15){\circle*{3}}
\put(112,21){$\pi$} \put(110,30){\circle*{3}}
\put(110,15){\circle{3}} \put(116,11){\framebox(8,8){$Not$}}
\put(130,30){\circle*{3}} \put(130,34){$\pi\over2$}
\put(130,15){\circle*{3}} \put(127,19){$-{\pi\over 2}$}
\put(135,10){\dashbox{2}(0,25)} \put(140,30){\circle*{3}}
\put(140,33){$\pi$} \put(160,30){\circle*{3}}
\put(160,19){\line(0,1){11}} \put(147,30){\circle{3}}
\put(147,15){\line(0,1){15}} \put(147,15){\circle*{3}}
\put(156,11){\framebox(8,8){$U$}}
\put(180,26){\framebox(8,8){$H$}} \put(124,15){\line(1,0){32}}
\put(180,15){\line(1,0){20}} \put(164,15){\line(1,0){16}}
\end{picture}}
\end{equation}
Here the \textit{Not gate} denotes\footnote{In fact a \textit{Not gate} could also be built directly.}:
\begin{equation}
\mbox{\hspace{4,5cm}}
\mbox{
\begin{picture}(100,20)(50,15)
\put(0,16){\framebox(8,8){$Not$}}
\put(38,16){\framebox(8,8){$H$}}
\put(-5,20){\line(1,0){5}}
\put(8,20){\line(1,0){5}}
\put(20,20){=}
\put(53,20){\circle*{3}}
\put(60,16){\framebox(8,8){$H$}}
\put(33,20){\line(1,0){5}}
\put(46,20){\line(1,0){14}}
\put(53,22){$\pi$}
\put(68,20){\line(1,0){5}}
\end{picture}}
\end{equation}

Again it can be proved that this gate prepares the correct state
$\ket{\psi}_{AB}$ and projects onto $\ket{\phi_j}$ as given in
(\ref{seventeen}). The first part gives:
\begin{eqnarray}
     \vert\bold 0\rangle=\vert0\rangle\otimes\vert0\rangle\;\;\stackrel{Not_2}{\mapsto}\;\;\vert0\rangle\otimes\vert1\rangle\;\;\stackrel{CN_1}{\mapsto}\;\;\vert1\rangle\otimes\vert1\rangle\;\;\stackrel{H_2}{\mapsto}\;\;{1\over\sqrt2}(\vert10\rangle-\vert11\rangle)\;\;\stackrel{CN_1}{\mapsto}
      \;\;{1\over\sqrt2}(\vert10\rangle-\vert01\rangle).
\end{eqnarray}
To prove that the second part projects onto (\ref{seventeen}), we
first look at the first three steps in the diagram:
\begin{eqnarray}
 \vert\phi_{1,2}\rangle&=&{1\over\sqrt{2}}\vert01
    \rangle\pm{1\over 2}\left[\vert11\rangle\cdot
    e^{i\pi\over4}-\vert00\rangle\cdot
    e^{-i\pi\over4}\right]\nonumber\\
    &\stackrel{CP(\pi)}{\mapsto}&{1\over\sqrt2}\vert01\rangle\pm{1\over2}\left[-\vert11\rangle\cdot
    e^{i\pi\over4}-\vert00\rangle\cdot e^{-i\pi\over4}\right]\nonumber\\
    &\stackrel{Not_2}{\mapsto}&  {1\over\sqrt2}\vert00\rangle\pm{1\over2}\left[-\vert10\rangle\cdot
    e^{i\pi\over4}-\vert01\rangle\cdot e^{-i\pi\over4}\right]\nonumber\\
    &\stackrel{P({\pi\over2})_1,P(-{\pi\over2})_2}{\mapsto}&{1\over\sqrt{2}}\vert00
    \rangle\pm{1\over 2}\left[\vert01\rangle\cdot
    e^{i\pi\over4}+\vert10\rangle\cdot
    e^{-i\pi\over4}\right]\nonumber\\
    \vert\phi_{3,4}\rangle&=&{1\over\sqrt{2}}\vert10
    \rangle\pm{1\over 2}\left[\vert11\rangle\cdot
    e^{i\pi\over4}-\vert00\rangle\cdot
    e^{-i\pi\over4}\right] \nonumber\\
    &\stackrel{CP(\pi)}{\mapsto}&{1\over\sqrt2}\vert10\rangle\pm{1\over2}\left[-\vert11\rangle\cdot
    e^{i\pi\over4}-\vert00\rangle\cdot e^{-i\pi\over4}\right]\nonumber\\
    &\stackrel{Not_2}{\mapsto}&  {1\over\sqrt2}\vert11\rangle\pm{1\over2}\left[-\vert10\rangle\cdot
    e^{i\pi\over4}-\vert01\rangle\cdot e^{-i\pi\over4}\right]\nonumber\\
    &\stackrel{P({\pi\over2})_1,P(-{\pi\over2})_2}{\mapsto}&{1\over\sqrt{2}}\vert11
    \rangle\pm{1\over 2}\left[\vert01\rangle\cdot
    e^{i\pi\over4}+\vert10\rangle\cdot
    e^{-i\pi\over4}\right].
\end{eqnarray}
But this is the basis (\ref{sixteen}). Therefore, these gates
transform one basis into the other. The rest of the gate is the
same as in (\ref{qgate1}) and hence the gate projects onto the
correct basis.

\section{Appendix}

In the appendix we give some basic formulae as well as a different
basis for the $m$=3 orthogonal case and the basis for the $m$=3
non-orthogonal case.

(i), Basic Transformations:

\begin{eqnarray}
      \ket\uparrow_x&=&{1\over\sqrt2}\cdot(\ket\uparrow_z+\ket\downarrow_z)\nonumber\\
      \ket\downarrow_x&=&{1\over\sqrt2}\cdot(\ket\uparrow_z-\ket\downarrow_z)\nonumber\\
      \ket\uparrow_y&=&{1\over\sqrt2}\cdot(\ket\uparrow_z+i\ket\downarrow_z)\nonumber\\
      \ket\downarrow_y&=&{1\over\sqrt2}\cdot(\ket\uparrow_z-i\ket\downarrow_z).
\end{eqnarray}

(ii), In \cite{Aharanov} the state (\ref{fifteen}) was used and the following
basis was obtained:
\begin{eqnarray}
    \vert\phi_1\rangle&=&{1\over\sqrt{2}}\ket{\uparrow \uparrow}_z+{1\over 2}\left[\ket{ \uparrow \downarrow }_z\cdot
    e^{i\pi\over4}+\ket{ \downarrow \uparrow }_z\cdot e^{-{i\pi\over4}}\right]\nonumber\\
    \vert\phi_2\rangle&=&{1\over\sqrt{2}}\ket{\uparrow \uparrow}_z-{1\over 2}\left[\ket{ \uparrow \downarrow }_z\cdot
    e^{i\pi\over4}+\ket{\downarrow \uparrow}_z\cdot e^{-{i\pi\over4}}\right]\nonumber\\
    \vert\phi_3\rangle&=&{1\over\sqrt{2}}\ket{\downarrow\downarrow}_z+{1\over 2}\left[\ket{\uparrow \downarrow}_z\cdot
    e^{-{i\pi\over4}}+\ket{ \downarrow \uparrow }_z\cdot e^{i\pi\over4}\right]\nonumber\\
   \vert\phi_4\rangle&=&{1\over\sqrt{2}}\ket{ \downarrow\downarrow}_z-{1\over 2}\left[\ket{ \uparrow \downarrow }_z\cdot
   e^{-{i\pi\over4}}+\ket{\downarrow \uparrow }_z\cdot
    e^{i\pi\over4}\right]. \label{sixteen}
\end{eqnarray}
This is equivalent to (\ref{seventeen}) and can be obtained by a unitary transformation.

(iii), The $m=3$ non-orthogonal case:

In this case Alice can choose a look-up table like:

\bigskip
\begin{center}
\mbox{
\begin{tabular}{|c||c|c|c|} \hline
         &      $\vec n_1$   &     $\vec n_2$  &     $\vec n_3$    \\           \hline \hline
$\lambda_1$ & $\;\;\;\;\downarrow\;\;\;\;$ & $\;\;\;\;\uparrow\;\;\;\;$ & $\;\;\;\;\downarrow\;\;\;\;$  \\  \hline
$\lambda_2$ & $\downarrow$ & $\uparrow$ & $\uparrow$     \\ \hline
$\lambda_3$ & $\downarrow$ & $\downarrow$ & $\downarrow$ \\ \hline
$\lambda_4$ & $\downarrow$ & $\downarrow$ & $\uparrow$   \\ \hline
$\lambda_5$ & $\uparrow$   & $\uparrow$&$\downarrow$     \\ \hline
$\lambda_6$ & $\uparrow$   & $\downarrow$& $\uparrow$    \\ \hline
$\lambda_7$ & $\uparrow$   & $\uparrow$&$\uparrow$       \\ \hline
$\lambda_8$ & $\uparrow$   & $\uparrow$& $\uparrow$      \\ \hline
\end{tabular}}
\end{center}
\bigskip

A special choice of the orientation of axes with $b_1^2=b_4^2=b_5^2=b_6^2=b_7^2=b_8^2={1\over8}$ and $b_2=0,\;
b_3^2={1\over8}$ leads to:

\begin{eqnarray}
     \vert\phi_1\rangle&=&\ket{\downarrow\uparrow\downarrow}\nonumber\\
     \vert\phi_2\rangle&=&-{1\over4}\left(1+i\sqrt3\right)\ket{\uparrow\uparrow\uparrow}-{1\over4}\left(1-i\sqrt3\right)
     \vert\uparrow\uparrow\downarrow\rangle+{1\over4}\left(\ket{\uparrow\downarrow\uparrow}+\ket{\uparrow\downarrow\downarrow}\right)+{\sqrt{35}\over10}\ket{\downarrow\downarrow\uparrow}-{\sqrt{10}\over20}
     \ket{\downarrow\downarrow\downarrow}\nonumber\\
     \vert\phi_3\rangle&=&+{1\over4}\left(1+i\sqrt3\right)\ket{\uparrow\uparrow\uparrow}+{1\over4}\left(1-i\sqrt3\right)\
     \ket{\uparrow\uparrow\downarrow}+{1\over4}\left(\ket{\uparrow\downarrow\uparrow}+\ket{\uparrow\downarrow\downarrow}\right)+{\sqrt{35}\over10}\ket{\downarrow\downarrow\uparrow}-{\sqrt{10}\over20}\
     \ket{\downarrow\downarrow\downarrow}
     \nonumber\\
     \vert\phi_4\rangle&=&\left(-{\sqrt2\over4}+i{\sqrt6\over12}\right)\ket{\uparrow\uparrow\uparrow}+\left(-{\sqrt2\over4}
     -i{\sqrt(6)\over12}\right)\ket{\uparrow\uparrow\downarrow}+\left({\sqrt2\over4}-{\sqrt3\over6}\right)\ket{\uparrow\downarrow\uparrow}+
     \left({\sqrt2\over4}+{\sqrt3\over6}\right)\ket{\uparrow\downarrow\downarrow}+\nonumber\\
     &&+{\sqrt7\over7}\ket{\downarrow\uparrow\uparrow}-{1\over{35}}\ket{\downarrow\downarrow\uparrow}+{\sqrt5\over10}\ket{\downarrow\downarrow\downarrow} \nonumber\\
    \vert\phi_5\rangle&=&\left({1\over4}-i{\sqrt3\over12}\right)\ket{\uparrow\uparrow\uparrow}+\left({1\over4}+i{\sqrt3\over12}\right)\ket{\uparrow\uparrow\downarrow}+\left({1\over4}+{\sqrt6\over12}\right)\ket{\uparrow\downarrow\uparrow}+\left({1\over4}-{\sqrt6\over12}\right)\ket{\uparrow\downarrow\downarrow}+{\sqrt{10}\over4}\ket{\downarrow\downarrow\downarrow} \nonumber\\
     \vert\phi_6\rangle&=&
     \left({1\over4}-i{\sqrt3\over12}\right)\ket{\uparrow\uparrow\uparrow}+\left({1\over4}+i{\sqrt3\over12}\right)\ket{\uparrow\uparrow\downarrow}+\left({1\over4}+{\sqrt6\over12}\right)\ket{\uparrow\downarrow\uparrow}+\left({1\over4}-{\sqrt6\over12}\right)\ket{\uparrow\downarrow\downarrow}-{\sqrt{14}\over7}\ket{\downarrow\uparrow\uparrow}-\nonumber\\
     &&- {2\over 35}
     \sqrt{35}\ket{\downarrow\downarrow\uparrow}-{3\over{20}}\sqrt{10}\ket{\downarrow\downarrow\downarrow}\nonumber\\
     \vert\phi_7\rangle&=&\left({1\over4}-i{\sqrt3\over12}\right)\ket{\uparrow\uparrow\uparrow}+\left({1\over4}+i{\sqrt3\over12}\right)\ket{\uparrow\uparrow
     \downarrow}+\left({1\over4}+{\sqrt6\over6}\right)\ket{\uparrow\downarrow\uparrow}+\left({1\over4}-{\sqrt6\over6}\right)\ket{\uparrow\downarrow\downarrow}-{\sqrt{14}\over7}\ket{\downarrow\uparrow\uparrow}-\nonumber\\
     &&-{3\over70}\sqrt{35}\ket{\downarrow\downarrow\uparrow}-{\sqrt{10}\over{20}}\ket{\downarrow\downarrow\downarrow}\nonumber\\
     \vert\phi_8\rangle&=&\left(-{1\over4}+i{\sqrt3\over12}\right)\ket{\uparrow\uparrow\uparrow}+\left(-{1\over4}-i{\sqrt3\over12}\right)\ket{\uparrow\uparrow\downarrow}+\left({1\over4}+{\sqrt6\over6}\right)\ket{\uparrow\downarrow\uparrow}+\left({1\over4}-{\sqrt6\over6}\right)\ket{\uparrow\downarrow\downarrow}-{\sqrt{14}\over7}\ket{\downarrow\uparrow\uparrow}-\nonumber\\
     &&-{3\over70}\sqrt{35}\ket{\downarrow\downarrow\uparrow}-{\sqrt{10}\over{20}}\ket{\downarrow\downarrow\downarrow}, \nonumber\\
\end{eqnarray}
all expressed in the z-basis. As in (\ref{unitary}), all
parameters of the unitary transformation were chosen to be zero.

\section{Acknowledgements}

I would like to thank Prof. Artur Ekert and Daniel Oi for their
assistance and support as well as for very many interesting
discussions. Furthermore, I would like to thank the authors of
\cite{Ekert} for permission to use the diagrams in (\ref{Phase
Shift}), (\ref{Hadamard
  Gate}), (\ref{Controlled Not}) and (\ref{Controlled Phase Shift}).


\begin{thebibliography}{99}

\bibitem{Aharanov}
L.Vaidman, Y.Aharanov, D.Z.Albert, {\em Phys. Rev. Lett.} {\bf 58} 1385 (1987).

\bibitem{Ben-Menahem}
S.Ben-Menahem, {\em Phys. Rev. A} {\bf 39} 1621 (1988).

\bibitem{Bub}
J.Bub, "Secure Key Distribution via Pre- and Post-Selected
Quantum Ensembles", e-print quant-ph/0006086, (2000).

\bibitem{IBM}
A.Barenco, C.H.Bennett, R. Cleve, D.P.DiVincenzo, N.Margolus, P.Shor, T.Sleator, J.Smolin, H.Weinfurter, {\em
  Phys. Rev. A} {\bf 52} 3457-3467 (1995).

\bibitem{Ekert}
A.Ekert, M.Ericsson, P.Hayden, H.Inamori, J.A.Jones, D.K.L.Oi,
V.Vedral, "Geometric Quantum Computation", e-print
quant-ph/0004015, (2000).



\end{thebibliography}
\end{document}